\documentclass[aps,prd,twocolumn,groupedaddress,showpacs,nofootinbib,amssymb,balancelastpage]{revtex4}
\usepackage[dvipdfmx]{color}
\usepackage[dvipdfmx]{graphicx}
\usepackage{bm}
\usepackage{amsmath}
\usepackage{amssymb}
\usepackage{amsfonts}
\usepackage{cases}
\usepackage{cancel}

\begin{document}

\title{Dark Matter in Modified Gravity?}
\author{Taishi Katsuragawa, Shinya Matsuzaki}
\affiliation{
Department of Physics, Nagoya University, Nagoya 464-8602, Japan \\
}

\begin{abstract}
We explore a new horizon of modified gravity from the viewpoint of the particle physics.
As a concrete example, we take the $F(R)$ gravity to raise a question:
can a scalar particle (``scalaron'') derived from the $F(R)$ gravity be a dark matter candidate?
We place the limit on the parameter in a class of $F(R)$ gravity model from the constraint on the scalaron as a dark matter.
The role of the screening mechanism and compatibility with the dark energy problem are addressed.
\end{abstract}


\maketitle

\section{Introduction}
%There are four conventionally accepted fundamental interactions in our Universe: 
%Electromagnetic, weak, and strong interactions and gravity.
%They are considered to be described by the particle physics based on the Standard Model (SM) and the general relativity (GR),
%which have been tested and confirmed by many experiments and observations.
%Despite plenty of success in SM and GR, several phenomena are still unrevealed and mysterious.
%We have not identified the origin of Dark Energy (DE) and Dark Matter (DM).

Late-time accelerated expansion of the Universe has been confirmed by several independent observations
\cite{Perlmutter:1998np,Riess:1998cb,Spergel:2003cb,Spergel:2006hy,Komatsu:2008hk,Komatsu:2010fb,Tegmark:2003ud,Seljak:2004xh,Eisenstein:2005su}.
In order to explain the accelerated expansion, it could be inevitable to include the Dark Energy (DE) as the new energy source.
In addition to the existence of the DE, 
the observations of the rotation curve of galaxies and the gravitational lensing indicate the presence of new matters, 
which do not have the electromagnetic interaction but the gravitational one, so-called the Dark Matter (DM).
The origin of such a dark sector involving the DE and DM is still mysterious among fields of the particle physics and astrophysics.

The $\Lambda$CDM model provides the simplest way to account for the DE as well as the DM,
in which the cosmological constant $\Lambda$ and the Cold Dark Matter (CDM) are introduced in the framework of the general relativity (GR).
%The simplest way to account for the DE is to introduce a cosmological constant $\Lambda$ to the general relativity (GR).
%The well-known $\Lambda$CDM model is constructed, which successfully describes the almost all history of the Universe.
This model successfully describes the almost all of the cosmic history.
However, it suffers from several theoretical problems:
(i) the extremely large discrepancy between the theoretical and observational values of the cosmological constant, 
which is known as a fine-tuning problem;
(ii) the ratio of the DE to the CDM with respect to the current energy density,
which is known as a coincidence problem.
The $\Lambda$CDM does not give us any answer to these two problems, which would imply the necessity of a new paradigm. 

It would be the modified gravity theory that can solve those problems, which has been intensively investigated so far.
The modification of gravity brings a new degree of freedom, such as a scalar field, which can mimic the role of the cosmological constant
(for a review, see
\cite{Nojiri:2006ri,Nojiri:2010wj,Capozziello:2010zz,Bamba:2012cp,Joyce:2014kja,Koyama:2015vza,Clifton:2011jh,Capozziello:2011et}),
so that one can explain the late-time cosmic acceleration without invoking ad hoc introduction of the cosmological constant.
Hence, the problem (i) is initially not present.
(Note that, however, the fine-tuning problem of cosmological constant is translated into the fine-tuning of parameter in the modified gravity.)
Besides the DE problem,
it has been suggested that a new particle derived from the modification of gravity
can be a dark matter candidate
\cite{Nojiri:2008nk,Nojiri:2008nt,Katsuragawa:2013lfa}.
In this scenario, one might be able to predict the ratio of the DE to the DM regarding the energy density today, and give the answer to the problem (ii).

Of interest is that the DM candidate is naturally introduced merely due to the modification of gravitational theory, without any ad hoc assumptions.
More remarkably, this DM possesses a salient feature which cannot be seen either in the Weakly Interacting Massive Particle scenario, 
or in the axion DM scenario.
That is the screening mechanism:
the propagation of the DM can be regarded as the fifth force in the astrophysical observation, which is suppressed in the modified gravity.
In a similar way, the propagation of the DM in the Standard Model (SM) bulk should be suppressed. 
Thus, the screening mechanism would trigger the non-trivial effect on the coupling between the DM and the SM particles.
Recently, the DM from the modified gravity theory has been discussed in the context of particle physics
\cite{Choudhury:2015zlc,Aoki:2016zgp,Babichev:2016hir,Babichev:2016bxi}.
However, the screening mechanism was not taken into account when the DM couplings with the SM are formulated.

In this paper, we study a DM candidate derived from the $F(R)$ gravity, one of modified gravity theories, taking into account the screening mechanism properly.
The $F(R)$ gravity can be expressed as the scalar-tensor theory,
which includes the Einstein--Hilbert action and an extra scalar field.
This scalar is the DM candidate, which we shall call ``scalaron''.
If we regard the scalaron as a DM, it plays two different roles:
it acts as the DE at cosmological scale
while it is a DM candidate at smaller scales.
This scale-dependent behavior is reflected by the screening mechanism, which is called the chameleon mechanism in the $F(R)$ gravity.
Due to the chameleon mechanism,
the scalaron becomes heavy around the high density region, and the $F(R)$ gravity can avoid the Solar System constraint.

The key idea we shall propose is that
the chameleon mechanism is applied to the microscopic environment:
one naively expects that the scalaron becomes heavy even in the SM bulk
because the density of the ensemble made of the SM particles,
estimated from the macroscopic view, is high enough.
However, {\it it is nontrivial how the chameleon mechanism works for the scalaron
interacting with the SM in the framework of the particle physics},
although the chameleon mechanism is well understood on the astrophysical ground.

Based on this idea,
we derive the couplings between the scalaron and the SM particles in the quantum field theory.
We evaluate the lifetime of the scalaron by taking into account the chameleon mechanism,
to find the upper limit of the scalaron mass.
We then discuss the constraint on the form of the $F(R)$ function by the relation between the $F(R)$ function and the scalaron potential.

\section{$F(R)$ gravity and the Weyl transformation} \label{Short review of $F(R)$ gravity}

In this section, we give a brief review of the $F(R)$ gravity, 
and observe how the scalar field emerges via the Weyl transformation of the metric.
We begin with the action of $F(R)$ gravity defined as follows:
\begin{align}
S = \frac{1}{2\kappa^{2}} \int d^{4}x \sqrt{-g} F(R) + S_{\mathrm{Matter}} \, , 
\label{F(R)action}
\end{align}
where $\kappa^{2} = 8 \pi G = 1/M^{2}_{\mathrm{pl}}$ and $M_{\mathrm{pl}}$ is the reduced Planck mass.
$F(R)$ is a function of the Ricci scalar $R$: e.g. $F(R) = R$ in general relativity.
The matter action $S_{\mathrm{Matter}}$ is defined as
\begin{align}
S_{\mathrm{Matter}} = \int d^{4}x \sqrt{-g} \mathcal{L}_{\mathrm{Matter}} (g^{\mu \nu}, \Psi)
\label{MatterLagrangian}
\, .
\end{align}
Here, $\mathcal{L}_{\mathrm{Matter}}$ is the matter Lagrangian density and $\Psi$ denotes the matter fields.
From the action (\ref{F(R)action}), we obtain the following equation of motion,
\begin{align}
\kappa^{2} T_{\mu \nu}
=&
F_{R}(R)  R_{\mu \nu} 
- \frac{1}{2} g_{\mu \nu} F(R) 
\nonumber \\
& \quad
+ \left( g_{\mu \nu} \Box  
-  \partial_{\mu} \partial_{\nu} \right) F_{R}(R)  \, ,
\label{F(R)eom}
\end{align}
where $F_{R} = \partial_{R}F(R)$ for convenience.
And, the energy-momentum tensor $T_{\mu \nu}$ is defined as
\begin{align}
\delta \mathcal{L}_{\mathrm{Matter}} \equiv \frac{1}{2} \sqrt{-g} T_{\mu \nu} \delta g^{\mu \nu} \, .
\end{align}

Sharp contrast to the Einstein equation stems from the trace part of Eq.~(\ref{F(R)eom}):
by performing the trace of Eq.~(\ref{F(R)eom}), we obtain the Klein-Gordon type equation,
\begin{align}
\Box F_{R}(R) = \frac{1}{3}\kappa^{2} T + \frac{1}{3} \left[ 2F(R) - F_{R}(R) R \right] .
\label{F(R)eom_trace}
\end{align}
Eq.~(\ref{F(R)eom_trace}) implies that $R$ is dynamical 
although it is determined by the algebraic relation with the energy-momentum tensor, $R=-\kappa^{2}T$, in the general relativity.
This scalar degree of freedom corresponds to the scalar function $F(R)$,
and it appears if and only if $F(R) \neq R$.

Next, we extract the dynamics of the scalar degree of freedom in the $F(R)$ gravity.
We first rewrite the action (\ref{F(R)action}) by introducing  an auxiliary field $A$ in the following form:
\begin{align}
S = \frac{1}{2\kappa^{2}} \int d^{4}x \sqrt{-g} \left[ F_{A}(A)R  - \left \{ F_{A}(A) A  - F(A) \right \} \right] \, .  
\label{F(R)action2}
\end{align}
By the variation of (\ref{F(R)action2}) with respect to $A$, we obtain the equation of motion for $A$: $F_{AA} (A) \left( R- A \right) = 0 $,
and we find $A=R$ if $F_{RR} (R) \neq 0$ for all $R$.
Substituting $A=R$ into the action (\ref{F(R)action2}) again, one can reproduce the original action (\ref{F(R)action}).

The above relation between $A$ and $R$ is consistent with the fact that the Ricci scalar is dynamical in the $F(R)$ gravity as Eq.~(\ref{F(R)eom_trace}) implies.
Therefore, we find that the dynamics of $F(R)$ gravity is equivalent to the dynamics of the general relativity 
with the non-minimal coupling to the scalar field $A$.
Note that this newly introduced scalar field brings a modification of gravity, 
and plays a significant role in cosmology if the mass of the scalar field is as small as the cosmological constant.

Next, we deform the non-minimal coupling between the scalar field $A$ and the Ricci scalar $R$ into the minimal coupling by the Weyl transformation.
The Weyl transformation of the metric is defined as $g_{\mu \nu} \rightarrow \tilde{g}_{\mu \nu} = \mathrm{e}^{2 \sigma(x)}g_{\mu \nu}$.
It can be seen as the transformation of frame; 
the Jordan frame described by the original metric $g_{\mu \nu}$ is transformed to the Einstein frame with $\tilde{g}_{\mu \nu}$.
Under the Weyl transformation, the line element in the Einstein frame is
\begin{align}
d\tilde{s}^{2} 
=& \tilde{g}_{\mu \nu} d\tilde{x}^{\mu}d\tilde{x}^{\nu} \nonumber \\
=&  \mathrm{e}^{2 \sigma} g_{\mu \nu} dx^{\mu}dx^{\nu} 
=  \mathrm{e}^{2 \sigma} ds^{2} \, .
\label{line_element}
\end{align}
Here, we note that 
the Weyl transformation changes the distances between the two points described by the same coordinate system $x^{\mu}$ on the manifold.
Thus, the definitions of time and length are different between two frames:
after the calculation in the Einstein frame,
dimensionful observables in the Jordan frame are evaluated by the scale transformation according to (\ref{line_element}).
Hereafter, we will use the partial derivative as $\partial_{\mu}$ (with lower index) and the coordinate as $x^{\mu}$ (with upper index),
to avoid confusion in raising and lowering the indices by $g_{\mu \nu}$ or $\tilde{g}_{\mu \nu}$.

Under the Weyl transformation, the action (\ref{F(R)action2}) is transformed into the following form:
\begin{align}
S 
=& 
\frac{1}{2\kappa^{2}} \int d^{4}x \sqrt{-\tilde{g}} 
\left[ F_{A}(A) \mathrm{e}^{-2\sigma} 
\right.
\nonumber \\
& \quad \times
\left\{ 
\tilde{R}
+ 6 \tilde{g}^{\mu \nu} \tilde{\nabla}_{\mu} \partial_{\nu} \sigma 
- 6 \tilde{g}^{\mu \nu} \left( \partial_{\mu} \sigma \right) \left( \partial_{\nu} \sigma \right) 
\right\} 
\nonumber \\
&  \left.
\quad
- \mathrm{e}^{-4\sigma} \left \{ F_{A}(A) A  - F(A) \right \} \right]  \, .
\label{F(R)action3}
\end{align}
By defining the Weyl transformation as $\mathrm{e}^{2\sigma} \equiv F_{A}(A)$,
the non-minimal coupling between the Einstein--Hilbert action and the scalar field vanishes
in the action (\ref{F(R)action3}).
Furthermore, by redefining the variable as $\varphi(x) \equiv\sqrt{6} \sigma(x)/ \kappa$,
the kinetic term of the scalar field is canonically normalized.
We then find the $F(R)$ gravity is expressed as the general relativity minimally coupling with the scalar field:
\begin{align}
S
=&
\frac{1}{2\kappa^{2}} \int d^{4}x \sqrt{-\tilde{g}}   
\tilde{R}
+ \int d^{4}x \sqrt{-\tilde{g}} 
\left[ 
- \frac{1}{2} \tilde{g}^{\mu \nu} \left( \partial_{\mu} \varphi \right) \left( \partial_{\nu} \varphi \right) 
\right. 
\nonumber \\
& \left. \quad
- \frac{1}{2\kappa^{2}} \frac{ F_{A}(A(\varphi)) A (\varphi) - F(A(\varphi))  }{F_{A}^{2}(A(\varphi))}
\right] \, .
\label{F(R)action5}
\end{align}
The gravitational theory described as in the action (\ref{F(R)action5}) 
is called the scalar-tensor theory:
the scalar field $A$ acts as the gravitational force besides the tensor field $g_{\mu \nu}$.

Finally, we consider the effect of the scalar field to the matter sector.
According to the Weyl transformation, 
the matter action (\ref{MatterLagrangian}) is expressed as
\begin{align}
S_{\mathrm{Matter}} 
=& 
\int d^{4}x \sqrt{-\tilde{g}} \, \mathrm{e}^{-4\sqrt{1/6}\kappa \varphi} \nonumber \\
& \times \mathcal{L}_{\mathrm{Matter}} \left( g^{\mu \nu}= \mathrm{e}^{2\sqrt{1/6}\kappa \varphi} \tilde{g}^{\mu \nu}, \Psi \right) \, .
\label{ScalaronMatterCoupling}
\end{align}
After the Weyl transformation, 
the ``dilatonic'' coupling of the scalar field $\varphi$ to the matter fields $\Psi$ shows up
in (\ref{ScalaronMatterCoupling}).
Note that the interactions between the scalaron and SM particles are derived not only from the Lagrangian,
but also from $\sqrt{-g}$.
This scalar field propagates between the matter fields besides the graviton.
Hereafter, we refer to this scalar field $\varphi$ as ``scalaron'' in order to distinguish from the other matter fields.

\section{$F(R)$ gravity for the Dark Energy}

As we saw in the previous section,
the modification of gravitational action leads to a new degree of freedom,
and its couplings to the ordinary matter field are necessarily introduced.
Then, the new degree of freedom modifies the gravitational interaction,
which causes the different prediction from the general relativity.
On the other hand, in order to be a gravitational theory, 
the modified gravity should explain or satisfy the astrophysical observations and constraints.
In this section, we give a brief review about the requirements to avoid the Solar System constraint.

\subsection{Scalaron potential and mass}

Constraints from the violation of the equivalence principle in the Solar System 
often exclude modifications of gravity
although the modifications are required for the DE in the cosmological scale.
Thus, we need to suppress the fifth force mediated by the new degree of freedom only in the smaller scale.
Viable models of $F(R)$ gravity have a screening mechanism to screen the fifth force 
mediated by the scalaron and avoid the constraint,
which is called the chameleon mechanism.
In this subsection, we review the chameleon mechanism in the $F(R)$ gravity.

We first consider the equation of motion for the scalaron field.
By variation of (\ref{F(R)action5}) with respect to the scalaron field $\varphi$, we obtain the equation of motion of the scalaron 
\begin{align}
\tilde{\Box} \varphi 
= V_{\mathrm{eff}}^{\prime} (\varphi) \, , 
\nonumber
\end{align}
where the effective potential $V_{\mathrm{eff}}(\varphi)$ is defined as
\begin{align}
V_{\mathrm{eff}}^{\prime}(\varphi) 
\equiv& V^{\prime} (\varphi) + \frac{\kappa}{\sqrt{6}}  \mathrm{e}^{-4\sqrt{1/6}\kappa \varphi} T^{\mu}_{\ \mu} 
\label{EffectivePotential}
\, .
\end{align}
We find that the scalaron couples to the trace of the energy-momentum tensor 
in the equation of motion.
For simplicity, we consider the non-relativistic perfect fluid with the constant energy density in the Jordan frame. 
The energy-momentum tensor is then expressed as
$T_{\mu \nu} = \mathrm{diag} \left[ \rho, 0, 0, 0 \right]$,
and $T^{\mu}_{\ \mu} = g^{\mu \nu} T_{\mu \nu} = -\rho$.
In this case, the effective potential (\ref{EffectivePotential}) is given by
\begin{align}
V_{\mathrm{eff}}(\varphi) =&
V(\varphi) + \frac{1}{4}  \mathrm{e}^{-4\sqrt{1/6}\kappa \varphi} \rho \, .
\label{ChameleonPotential}
\end{align}
If $V'(\varphi) > 0$, the effective potential $V_{\mathrm{eff}}(\varphi)$ has a minimum at $\varphi = \varphi_{\min}$,
which satisfies $V^{\prime}_{\mathrm{eff}}(\varphi_{\min}) = 0$ and $V^{\prime \prime}_{\mathrm{eff}}(\varphi_{\min}) > 0$.
Performing the Tayler expansion of the effective potential around $\varphi=\varphi_{min}$, 
we find
\begin{align}
V_{\mathrm{eff}}(\varphi) = V_{\mathrm{eff}}(\varphi_{\min}) 
+ \frac{1}{2} V^{\prime \prime}_{\mathrm{eff}}(\varphi_{\min}) \left(\varphi - \varphi_{\min}\right)^{2} + \cdots \, .
\label{EffectivePotential2}
\end{align}
Here, the mass of the scalaron field is defined 
as the coefficient of $\left(\varphi - \varphi_{min}\right)^{2}$ in Eq.~(\ref{EffectivePotential2}),
and we find
\begin{align}
m^{2}_{\varphi} 
\equiv& V^{\prime \prime}_{\mathrm{eff}}(\varphi_{\min})
= V^{\prime \prime} (\varphi_{\min}) 
+ \frac{2\kappa^{2}}{3} \, \mathrm{e}^{- 4\sqrt{1/6}\kappa \varphi_{\min}} \rho \, .
\label{ChameleonMass}
\end{align}

Therefore, if $\rho$ is larger, the scalaron becomes heavier:
in the bulk of the Universe, where the energy density is very small, 
the scalaron can be very light and produce the effective cosmological constant.
On the other hand, in or around the heavy objects, the Solar System or the Earth, 
the scalaron becomes heavy.
Then, the Compton wavelength becomes short and the scalaron is screened.

\subsection{Starobinsky model}

In the previous subsection,
we discussed the chameleon mechanism in the $F(R)$ gravity without specifying the function of $F(R)$.
In this subsection, 
we consider, for example, the Starobinsky model \cite{Starobinsky:2007hu}
for the late-time acceleration,
and see how the chameleon mechanism works.
The action of the Starobinsky model is defined as
\begin{align}
F(R) = R - \beta R_{c} \left[ 1 - \left( 1 + \frac{R^{2}}{R^{2}_{c}} \right)^{-n} \right]
\nonumber 
\end{align}
with constants $n$, $\beta$, and $R_{c}>0$.
$R_{c}$ is constant curvature in the Starobinsky model, which is comparable to the cosmological constant $R_{c} \sim \Lambda$.
In the limit $R/R_{c} \gg 1$, we find
\begin{align}
F(R) 
\approx& 
R  - \beta R_{c}  + \beta R_{c}  \left(\frac{R}{R_{c}} \right)^{-2n}. 
\label{Starobinsky2}
\end{align}
If we ignore the third term in the large curvature regime, 
the Starobinsky model restores the GR with the cosmological constant, $\Lambda = \beta R_{c}/2$.
In this subsection, we study the approximated model (\ref{Starobinsky2})
because the chameleon mechanism works in the large curvature regime $R\gg R_{c}$.

First, we calculate the scalaron potential.
From the definition of the Weyl transformation, we find 
$\mathrm{e}^{2\sqrt{1/6}\kappa \varphi} = 1 -2n \beta \left(R/R_{c} \right)^{-(2n+1)} $.
Here, we note that the scalaron field $\varphi$ is negative, 
and it goes to zero $\varphi \rightarrow 00$ as the curvature $R$ increases.
Assuming the curvature is larger than the cosmological constant, $R \gg R_{0}\sim R_{c}$, 
we find
\begin{align}
|\kappa \varphi| 
=  \frac{\sqrt{6}}{2} \left| 
\ln \left( 1 -2n \beta \left( \frac{R}{R_{c}} \right)^{-(2n+1)}  \right)
\right| \ll 1 \, .
\label{WeakCoupling}
\end{align}
Here, we note that $ 2n\beta \left( R/R_{c} \right)^{-(2n+1)}<1$ because $F_{R}(R)>0$:
for the consistency with the observation,
it is required for $F(R)$ gravity models to avoid the anti-gravity.
If $F_{R}(R)<0$, the coefficient in front of the Einstein-Hilbert part in (\ref{F(R)action3}) is negative.
In other words, the gravitational constant becomes negative, which leads to the anti-gravity.

When we consider the non-relativistic matter for $T_{\mu \nu}$,
the scalaron potential is given by
\begin{align}
V_{\mathrm{eff}}(\varphi)
=&
\frac{\beta R_{c}}{2\kappa^{2}} 
\mathrm{e}^{-4\sqrt{1/6}\kappa \varphi}
\nonumber \\
& \times
\left[ 1 - (2n+1) 
\left\{ \frac{1}{2n \beta} 
\left( 1 - \mathrm{e}^{2\sqrt{1/6}\kappa \varphi} \right) \right\}^{\frac{2n}{2n+1}} 
\right]
\nonumber \\
& \qquad
+ \frac{1}{4} \mathrm{e}^{-4\sqrt{1/6}\kappa \varphi} \rho 
\label{StarobinskyPotential0}
\, .
\end{align}
Here, we note that the chameleon mechanism works at the high density region
where the curvature should be large.
So, in the following calculation, we assume $|\kappa \varphi| \ll 1$ as in (\ref{WeakCoupling}).
In this limit, the effective potential (\ref{StarobinskyPotential0}) is approximated to be
\begin{align}
V_{\mathrm{eff}}(\varphi)
\approx&
\frac{\beta R_{c}}{2\kappa^{2}} 
\mathrm{e}^{-4\sqrt{1/6}\kappa \varphi}
\nonumber \\
& \times
\left[ 1 - (2n+1) 
\left( - \frac{\kappa \varphi}{\sqrt{6}n \beta} \right)^{\frac{2n}{2n+1}} 
+ \frac{\kappa^{2}\rho }{2\beta R_{c}}
\right] \, .
\label{StarobinskyPotential}
\end{align}

Second, we study the minimum of the effective potential.
The derivative of the effective potential (\ref{StarobinskyPotential}) is
\begin{align}
V^{\prime}_{\mathrm{eff}}(\varphi)
\approx&
- \frac{4\kappa}{\sqrt{6}}
\frac{\beta R_{c}}{2\kappa^{2}} 
\mathrm{e}^{-4\sqrt{1/6}\kappa \varphi}
\nonumber \\
& \times
\left[ 1 - (2n+1) 
\left( - \frac{\kappa \varphi}{\sqrt{6}n \beta} \right)^{\frac{2n}{2n+1}} 
\right.
\nonumber \\
& \left. \qquad
- \frac{1}{2\beta} 
\left( - \frac{\kappa \varphi}{\sqrt{6}n \beta} \right)^{-\frac{1}{2n+1}} 
+ \frac{\kappa^{2}\rho }{2\beta R_{c}}
\right] \, .
\label{StarobinskyPotential1}
\end{align}
Solving $V^{\prime}_{\mathrm{eff}}=0$ in (\ref{StarobinskyPotential1}), 
we find the minimum $\varphi = \varphi_{\min}$,
\begin{align}
\kappa \varphi_{\min}
=
- \sqrt{6}n \beta \left( \frac{R_{c}}{\kappa^{2}\rho } \right)^{2n+1} \, .
\label{StarobinskyMass}
\end{align}

Finally, we evaluate the scalaron mass.
The second derivative of the effective potential (\ref{StarobinskyPotential}) is calculated as
\begin{widetext}
\begin{align}
V^{\prime \prime}_{\mathrm{eff}}(\varphi)
\approx&
 \left( \frac{4\kappa}{\sqrt{6}} \right)^{2}
\frac{\beta R_{c}}{2\kappa^{2}} 
\mathrm{e}^{-4\sqrt{1/6}\kappa \varphi}
\nonumber \\
& \times
\left[ 1 - (2n+1) 
\left( - \frac{\kappa \varphi}{\sqrt{6}n \beta} \right)^{\frac{2n}{2n+1}}
+ \frac{\kappa^{2}\rho}{2\beta R_{c}}
- \frac{1}{\beta} 
\left( - \frac{\kappa \varphi}{\sqrt{6}n \beta} \right)^{-\frac{1}{2n+1}}
+  \frac{1}{8n(2n+1)\beta^{2} } 
\left( - \frac{\kappa \varphi}{\sqrt{6}n \beta} \right)^{-\frac{1}{2n+1}-1}
\right] \, .
\label{StarobinskyPotential2}
\end{align}
\end{widetext}
Substituting Eq.~(\ref{StarobinskyMass}) into Eq.~(\ref{StarobinskyPotential2}) with $\varphi = \varphi_{\min}$, 
we obtain the expression of the scalaron mass
\begin{align}
m^{2}_{\varphi}
\approx&
\frac{R_{c}}{6n(2n+1)\beta } 
\left( \frac{\kappa^{2}\rho}{R_{c}} \right)^{2(n+1)} \, .
\label{StarobinskyMass2}
\end{align}
We note that 
the scalaron mass depends on the energy density,
and increases like a power function of $\rho$.

\section{$F(R)$ gravity for Dark Matter: Scalaron as Dark matter} 

We have seen how the chameleon mechanism works in the $F(R)$ gravity theory.
In this section, we reconsider 
the new coupling between the scalaron $\varphi$ and the SM particles in the Einstein frame.
It may bring us a fascinating fact 
that the new massive scalar field is naturally derived from the modification of gravity 
without extending the particle contents of the SM.
In other words,
the modification of gravitational theory affects the particle physics,
and the phenomena beyond the SM may show up.

We recall the properties of the scalaron as we have seen so far:
(1) after the Weyl transformation,
dilatonic interactions between the scalaron and the SM appear. 
The coupling is suppressed in the large curvature regime;
(2) the scalaron mass is very large in the large curvature regime 
because of the chameleon mechanism.
As a result, the fifth force, the propagation of the scalaron, is suppressed, 
and it is consistent with the observational constraint.
These two natures of the scalaron imply that
a massive field weakly coupled with SM particles emerges
in the Solar System, or around the Earth.
Therefore, this property suggests that the scalaron could be a CDM.
In other words, the ``Darkness" of DM is justified by the dilatonic coupling with the Planck mass suppression and the chameleon mechanism in this scenario.

In order to study the scalaron field as a DM candidate, 
it is necessary to formulate the couplings between the scalaron and the SM particles. 
In this section, 
we investigate the matter coupling in Eq.~(\ref{ScalaronMatterCoupling})
and determine the form and magnitude of couplings.

\subsection{Coupling to massless fields}
First, we consider the massless vector field $A_{\mu}$.
The Lagrangian density $\mathcal{L}_{V}$ is given by
\begin{align}
\mathcal{L}_{V} \left( g^{\mu \nu}, A_{\mu} \right)
=&
 - \frac{1}{4} \mathrm{e}^{4\sqrt{1/6}\kappa \varphi} 
\tilde{g}^{\alpha \mu} \tilde{g}^{\beta \nu} F_{\alpha \beta}F_{\mu \nu} 
\label{VectorWeyl0}
\, .
\end{align}
For simplicity, we consider an abelian gauge field.
The generalization to the curved space-time can be made
by replacing the partial derivative with the covariant derivative in the field strength $F_{\mu \nu}$.
However, we find that the field strength does not change
because the Christoffel symbols are symmetric, 
and $F_{\mu \nu} = \partial_{\mu} A_{\nu} - \partial_{\nu} A_{\mu}$ even in the curved space-time.
Thus, the field strength $F_{\mu \nu}$ is invariant under the Weyl transformation,
so that the action in the Einstein frame is given just by the replacing $g^{\mu \nu} \rightarrow \tilde{g}^{\mu \nu}$:
\begin{align}
S
=&
\int d^{4}x \sqrt{-\tilde{g}} 
\mathcal{L}_{V} \left( \tilde{g}^{\mu \nu}, A_{\mu} \right) \, ,
\label{VectorWeyl}
\end{align}
where the exponential factor $\mathrm{e}^{4\sqrt{1/6}\kappa \varphi}$ in (\ref{VectorWeyl0})
has been canceled with $\mathrm{e}^{-4\sqrt{1/6}\kappa \varphi}$ in (\ref{ScalaronMatterCoupling}).

We find that the direct coupling between the massless vector field $A_{\mu}$ and the scalaron $\varphi$ does not arise through the field strength.
The same argument is applicable even for the non-abelian gauge field.
However, as clearly discussed in Appendix.~\ref{appendix:1},
the scalaron couples to the massless vectors at the quantum level through the scale anomaly,
although it does not at the level of classical dynamics.

Second, we consider the massless fermion $\psi(x)$.
Unlike the case of bosonic fields, we need special treatment for fermion fields.
The Lagrangian density $\mathcal{L}_{F}$ in curved space-time is given by
\begin{align}
\mathcal{L}_{F} \left( \gamma^{\mu}, \psi \right)
= i \bar{\psi} (x) \gamma^{\mu} \nabla_{\mu} \psi (x) \, .
\label{FermionLagrangianDensity}
\end{align}
$\gamma^{\mu}$ is the generalized Dirac gamma matrix in curved space-time, defined as
$\gamma^{\mu} (x)  \equiv e_{a}^{\ \mu} (x)  \gamma^{a}$ 
, where $\left\{ \gamma^{\mu}, \, \gamma^{\nu} \right\} = 2 g^{\mu \nu}$.
And, the vierbein $e^{a}_{\mu}$ is related to the metric as
$g_{\mu \nu}(x) =  \eta_{ab} \, e^{a}_{\ \mu}(x) e^{b}_{\ \nu}(x)$ 
with Latin letters $a, b \cdots$ for Lorentz indices, and Greek letters $\mu, \nu \cdots$ for space-time indices.
The covariant derivative for the spinor is given by
\begin{align}
\nabla_{\mu} \psi (x)
=& 
\partial_{\mu}\psi (x) + \frac{1}{8} \omega_{\mu ab}(x) [ \gamma^{a}, \, \gamma^{b} ] \psi (x) 
\nonumber
\, ,
\end{align}
where $\omega_{\mu ab}(x)$ is called a spin connection, defined as
$ w_{\mu ab} (x) =  e_{a \nu} \left( \partial_{\mu}e_{b}^{\ \nu} + \Gamma^{\nu}_{\mu \rho} e_{b}^{\ \rho}\right)$.
Under the Weyl transformation $\tilde{g}_{\mu \nu}=\mathrm{e}^{2\sigma}g_{\mu \nu}$, 
the vierbein $e^{a}_{\ \mu}$ and the generalized Dirac gamma matrix $\gamma^{\mu}$ transform as
$\tilde{e}^{a}_{\ \mu} = \mathrm{e}^{\sigma} e^{a}_{\ \mu}$ and 
$ \tilde{\gamma}^{\mu} = \mathrm{e}^{-\sigma}  \gamma^{\mu}$, respectively.
Hence, the spin connection transforms as
$\omega_{\mu ab} 
= \tilde{\omega}_{\mu ab} - \left( \tilde{e}_{a \mu} \tilde{e}_{b}^{\ \lambda} - \tilde{e}_{b\mu} \tilde{e}_{a}^{\ \lambda} \right) \partial_{\lambda} \sigma$.

Using those transformation rules,
we find that the Lagrangian density (\ref{FermionLagrangianDensity}) in the Einstein frame is given by
\begin{align}
\mathcal{L}_{F} \left( \gamma^{\mu}, \psi \right)
=&
\mathrm{e}^{\sqrt{1/6}\kappa \varphi} i \bar{\psi} \tilde{\gamma}^{\mu} \tilde{\nabla}_{\mu} \psi
\nonumber \\
&
- \frac{3 i }{2} \sqrt{\frac{1}{6}}\kappa  \mathrm{e}^{\sqrt{1/6}\kappa \varphi} \left( \partial_{\mu} \varphi \right) \bar{\psi} \tilde{\gamma}^{\mu} \psi  \, ,
\nonumber 
\end{align}
where we used $\sigma = \sqrt{1/6} \kappa \varphi $.
Then, the action in the Einstein frame takes the following form:
\begin{align}
S
=&
\int dx^{4} \sqrt{-\tilde{g}} \, \mathrm{e}^{-4\sqrt{1/6}\kappa \varphi} 
\mathcal{L}_{F} \left( \gamma^{\mu}, \psi \right)
\nonumber \\
=&
\int d^{4}x \sqrt{-\tilde{g}} 
\left[ 
\mathrm{e}^{- 3\sqrt{1/6}\kappa \varphi} i \bar{\psi} \tilde{\gamma}^{\mu} \tilde{\nabla}_{\mu} \psi
\right.
\nonumber \\
& \left. \qquad \qquad \quad
- \frac{3 i }{2} \sqrt{\frac{1}{6}}\kappa \mathrm{e}^{ -3\sqrt{1/6}\kappa \varphi} \left( \partial_{\mu} \varphi \right) \bar{\psi} \tilde{\gamma}^{\mu} \psi  
\right] 
\label{FermiWeyl}
\, .
\end{align}
Note that the couplings between the fermion field $\psi$ and scalaron $\varphi$ are generated 
in (\ref{FermiWeyl})
when $\kappa \varphi \neq 0$, in contrast to the case of massless vector fields.

One may transform the action (\ref{FermiWeyl}) into the canonical form
by redefining the fermion field $\psi \rightarrow \psi^{\prime} = \mathrm{e}^{-3/2 \sqrt{1/6} \kappa \varphi} \psi$,
to find that
the redefined massless fermion field does not couple with the scalaron because of the Weyl transformation invariance,
just like the case of massless vector field:
\begin{align}
S
=&
\int d^{4}x \sqrt{-\tilde{g}}  i \bar{\psi^{\prime}} \tilde{\gamma}^{\mu} \tilde{\nabla}_{\mu} \psi^{\prime} \, .
\end{align}
Thus, one can eliminate the scalaron coupling by the field redefinition in classical dynamics.

However, the scalaron would affect the quantum dynamics of fermion field.
The scalaron disappears from the action, but would be transfered in the path integral measure
because the field redefinition involves the scalaron field.
Actually, the modified path integral measure
induces the scale anomaly, then, the couplings between the scalaron and massless vector fields show up
(for detail, see Appendix.~\ref{appendix:1}).
Note also that this scale anomaly has nothing to do with the transformation between the Jordan and Einstein frames.

\subsection{Coupling to massive fields}

Finally, we consider the massive vector and fermion fields.
After the electroweak symmetry breaking,
the vector and fermion fields acquire the mass through the Higgs mechanism.
The mass term of the vector field $\mathcal{L}_{V-\mathrm{mass}}$ is given by
\begin{align}
\mathcal{L}_{V-\mathrm{mass}} \left( g^{\mu \nu}, A_{\mu} \right)
=&
- \frac{1}{2} m^{2}_{V} \mathrm{e}^{2\sqrt{1/6}\kappa \varphi} \tilde{g}^{\mu \nu} A_{\mu} A_{\nu} 
\nonumber
\, ,
\end{align}
where $m_{V}$ is the mass of the massive vector field.
So, the action in the Einstein frame is given by
\begin{align}
S
=&
\int dx^{4} \sqrt{-\tilde{g}} \, \mathrm{e}^{-4\sqrt{1/6}\kappa \varphi} 
\mathcal{L}_{V-\mathrm{mass}} \left( g^{\mu \nu}, A_{\mu} \right)
\nonumber \\
=&
\int dx^{4} \sqrt{-\tilde{g}} 
\left[ - \frac{1}{2} m^{2}_{V} \mathrm{e}^{-2\sqrt{1/6}\kappa \varphi} \tilde{g}^{\mu \nu} A_{\mu} A_{\nu} \right]
\label{VectorWeyl5}
\end{align}

Now, we divide the action (\ref{VectorWeyl5}) into two parts:
\begin{align}
S
=&
\int d^{4}x \sqrt{-\tilde{g}} 
\left[
\mathcal{L}_{V-\mathrm{mass}} \left( \tilde{g}^{\mu \nu}, A_{\mu} \right)
\right.
\nonumber \\
& \left. \qquad \qquad \qquad
+ \mathcal{L}_{V-\varphi} \left( \tilde{g}^{\mu \nu}, A_{\mu}, \varphi \right)
\right]
\label{VectorScalaron5}
\, ,
\end{align}
where 
\begin{align}
&\mathcal{L}_{V-\varphi} \left( \tilde{g}^{\mu \nu}, A_{\mu}, \varphi \right)
\nonumber \\
& \qquad \equiv
- \frac{1}{2} m^{2}_{V} \left( \mathrm{e}^{-2\sqrt{1/6}\kappa \varphi} - 1 \right)\tilde{g}^{\mu \nu} A_{\mu} A_{\nu} \, .
\label{VectorScalaron25}
\end{align}
The first term in (\ref{VectorScalaron5}) is the mass term of vector field in the Einstein frame 
where the metric is replaced as $g_{\mu \nu} \rightarrow \tilde{g}_{\mu \nu}$.
The second term describes the non-linear interaction between the massive vector field $A_{\mu}$ and scalaron $\varphi$.

As done in Eq.~(\ref{WeakCoupling}),
we consider the weak coupling limit $\kappa \varphi \ll1$, corresponding to the large curvature limit $R\gg R_{c}$.
In this limit, we can expand the dilatonic coupling $\mathrm{e}^{\kappa \varphi}$ in the Lagrangian density (\ref{VectorScalaron25})
with respect to $|\kappa \varphi|\ll1$, we find
\begin{align}
\mathcal{L}_{V-\varphi} \left( \tilde{g}^{\mu \nu}, A_{\mu}, \varphi \right)
=&
\frac{2\kappa \varphi }{\sqrt{6}} \cdot \frac{1}{2} m^{2}_{V} \tilde{g}^{\mu \nu} A_{\mu} A_{\nu} 
+ \mathcal{O}(\kappa^{2}\varphi^{2}) \, .
\label{VectorWeyl25}
\end{align}
Thus, we have the couplings between the massive vector field $A_{\mu}$ and the scalaron $\varphi$.

The mass term of the fermion field is given by
\begin{align}
\mathcal{L}_{F-\mathrm{mass}} \left( \psi \right)
= - m_{F} \bar{\psi} \psi 
\nonumber
\, ,
\end{align} 
where $m_{F}$ is the mass of the massive fermion field.
By redefining the field as in the case of massless fermion field,
one finds
\begin{align}
\mathcal{L}_{F-\mathrm{mass}} \left( \psi \right)
= - m_{F} \mathrm{e}^{3\sqrt{1/6}\kappa \varphi}  \bar{\psi^{\prime}} \psi^{\prime}
\nonumber
\, .
\end{align} 
As in the case of the massive vector field, we obtain
\begin{align}
S
=&
\int d^{4}x \sqrt{-\tilde{g}} 
\left[
\mathcal{L}_{F-\mathrm{mass}} \left( \psi^{\prime} \right)
+ \mathcal{L}_{F-\varphi} \left( \psi^{\prime}, \varphi \right)
\right] \, ,
\nonumber
\end{align}
where 
\begin{align}
\mathcal{L}_{F-\varphi} \left( \psi^{\prime}, \varphi \right)
\equiv
- m_{F} \left( \mathrm{e}^{-\sqrt{1/6}\kappa \varphi} - 1 \right) \bar{\psi^{\prime}} \psi^{\prime} \, .
\label{FermionScalaron2}
\end{align}
Expanding the Lagrangian density (\ref{FermionScalaron2}) with respect to $|\kappa \varphi|\ll1$, we find
\begin{align}
\mathcal{L}_{F-\varphi} \left( \psi^{\prime}, \varphi \right)
=&
\frac{\kappa \varphi }{\sqrt{6}}  \cdot m_{F} \bar{\psi^{\prime}} \psi^{\prime}  
+ \mathcal{O}(\kappa^{2}\varphi^{2}) \, . 
\label{FermionWeyl4}
\end{align}
Thus, we have the couplings between the massive fermion field $\psi^{\prime}$ and the scalaron $\varphi$.

\section{An effective model for scalaron particle}

In the previous section,
the interactions between the scalaron and the SM particles have been investigated.
However, we need to discuss the particle picture of the scalaron field
although we usually study the classical dynamics of the scalaron field as a gravitational theory.

\subsection{Particle picture of the scalaron}

We shall first expand the scalaron field $\varphi$ around the background solution $\varphi=\varphi_{\min}$:
$\varphi = \hat{\varphi} + \varphi_{\min}$, and treat the fluctuation $\hat{\varphi}$ as a particle.
We now consider the role of the chameleon mechanism for the background solution $\varphi_{\min}$ and the scalaron ``particle'' $\hat{\varphi}$, respectively.

As we discussed in Eq.~(\ref{ChameleonPotential}), the scalaron potential changes through the chameleon mechanism according to the trace of the energy-momentum tensor. 
Hence, the chameleon mechanism reflects the environment dependence of the scalaron:
the energy-momentum tensor consists of the matter fields surrounding the scalaron, which controls the environment system. 
In order to take the chameleon mechanism into account,
we need to specify the energy-momentum tensor corresponding to the SM environment.
In the present analysis, we assume that the SM bulk is described by the perfect fluid,
and the energy density $\rho$ is namely given as $\rho_{\mathrm{EW}} \sim (100 \, \mathrm{GeV} )^{4}$
in Eq.~(\ref{ChameleonPotential}).

Thus, the explicit environment dependence enters in the background solution $\varphi_{min}$ through the equation of motion,
\begin{align}
\hat{\Box} \varphi_{\min} 
=
V^{\prime}(\varphi_{\min})
- \frac{\kappa}{\sqrt{6}} \mathrm{e}^{-4\sqrt{1/6} \kappa \varphi_{\min}} \rho_{\mathrm{EW}}
\nonumber
\, .
\end{align}
As for the scalaron particle $\hat{\varphi}$, the environment dependence implicitly arises in the mass expression,
which is given by substituting $\rho = \rho_{\mathrm{EW}}$ into Eq.~(\ref{ChameleonMass}),
\begin{align}
m^{2}_{\hat{\varphi}}
= V^{\prime \prime} (\varphi_{\min}) 
+ \frac{2\kappa^{2}}{3} \, \mathrm{e}^{- 4\sqrt{1/6}\kappa \varphi_{\min}} \rho_{\mathrm{EW}} \, .
\label{ChameleonMass2}
\end{align}

Performing the fluctuation around the background solution,
we find 
\begin{align}
S_{\mathrm{Matter}} 
=& 
\int d^{4}x \sqrt{-\tilde{g}} 
\, \mathrm{e}^{-4\sqrt{1/6}\kappa \left(\varphi_{\min} + \hat{\varphi} \right)}
\nonumber \\
& \qquad \times
\mathcal{L}_{\mathrm{SM}}
\left( \mathrm{e}^{2\sqrt{1/6}\kappa \left(\varphi_{\min} + \hat{\varphi} \right)} \tilde{g}^{\mu \nu}, \Psi \right)
\nonumber
\, .
\end{align}
Because $|\kappa \varphi_{\min}| \ll 1$, the scalaron coupling to the SM is approximately given by
\begin{align}
S_{\mathrm{Matter}} 
\approx&
\int d^{4}x \sqrt{-\tilde{g}} \, \mathrm{e}^{-4\sqrt{1/6}\kappa \hat{\varphi}}
\nonumber \\
& \qquad \quad \times 
\mathcal{L}_{\mathrm{SM}} \left( \mathrm{e}^{2\sqrt{1/6}\kappa \hat{\varphi}} \tilde{g}^{\mu \nu}, \Psi \right) \, , 
\label{ScalaronCoupling}
\end{align}
so that we utilize the result in the previous section just
by replacing $\varphi \rightarrow \hat{\varphi}$.
The environment dependence still implicitly remains in the scalaron mass in (\ref{ChameleonMass2})
while the effect of background solution in the scalaron coupling has been ignored in (\ref{ScalaronCoupling}).
Note that
we are considering the microscopic environment 
where the scalaron is touching with the SM particles.
When one discusses the chameleon mechanism in the Solar System,
the averaged energy density $\rho_{\odot} \sim 10 \, \mathrm{g/cm^{3}} = 10^{19} \, \mathrm{eV^{4}}$,
which implies $\rho_{\mathrm{EW}}$ is large enough to cause the chameleon mechanism
even in the microscopic environment.

\subsection{Lifetime of scalaron and constraint to parameters in $F(R)$ models}

Because the scalaron potential depends on the function of $F(R)$ as well as the environment,
the constraint as a DM candidate can be rephrased as the constraint 
on the form of the $F(R)$ function.
In this subsection, we study the decay process of the scalaron, evaluate the lifetime, and give the constraint on the parameter in the Starobinsky model.

\begin{figure}[htbp]
\centering
\includegraphics[width=0.8\linewidth]{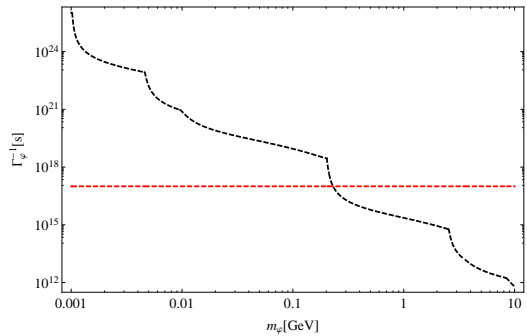} 
\caption{ 
The scalaron lifetime as a function of the mass in unit of GeV. 
The lifetime has been computed by summing up 
the partial decay widths in Eqs.(\ref{decay:FF}) and (\ref{widths:2}) relevant to 
the mass range displayed here. 
The horizontal (red) dashed line corresponds to the age of the universe $\simeq 10^{17}$s. } 
\label{plot-lifetime}
 \end{figure}

For the scalaron to be a dark matter candidate, 
the scalaron lifetime has to be longer than the age of the universe. 
As explicitly presented in Appendix~\ref{appendix:2},  
the scalaron can decay to the SM particles depending on 
its mass. Figure~\ref{plot-lifetime} shows 
the lifetime, inverse of the total decay width $\Gamma_\varphi$, as a fucntion of 
the scalaron mass. 
Up to the mass of 1 GeV scaling down from the higher mass, 
the scalaron dominantly decays to the tau lepton, charm quark pairs, 
so the scalaron with such a GeV mass cannot be present today.  
After the decay channels to tau and charm pairs get closed at the mass scale lower than 1 GeV,  
the scalaron still promptly decays to the strange quark and muon pairs. 
Thus, 
the upper bound on the scalaron mass is read off from Fig.~\ref{plot-lifetime} to be  
\begin{equation} 
 m_\varphi \lesssim 0.23 \,{\rm GeV} 
 \,. \label{MassBound} 
\end{equation}

Finally, we convert the mass bound (\ref{MassBound}) into the constraint on the $F(R)$ function.
For the Starobinsky model with $\beta R_{c} = 2\Lambda$, the scalaron mass is given by substituting $\rho = \rho_{\mathrm{EW}}$ into Eq.~(\ref{StarobinskyMass2}):
\begin{align}
m^{2}_{\hat{\varphi}} 
\approx \frac{2\Lambda}{6n(2n+1)\beta^{2} } 
\left( \frac{\kappa^{2} \beta}{2\Lambda} \rho_{\mathrm{EW}} \right)^{2(n+1)} 
\nonumber
\, ,
\end{align}
which leads to the constraint on the parameter $\beta$:
$\beta \lesssim 10^{-69}$ for $n=1$, and $\beta \lesssim 10^{-59}$ for $n=4$.
We note that the constraint obtained here has been estimated in the Einstein frame.
One would obtain almost the same result in magnitude even in the Jordan frame
because the Weyl transformation operates as almost identity
$\mathrm{e}^{2\sigma} \sim 1$, i.e., $\tilde{g}_{\mu \nu} \sim g_{\mu \nu}$.

\section{Summary and Discussion}

We have studied the scalaron field in the $F(R)$ gravity from the viewpoint of particle physics.
We have assumed that the scalaron is a DM candidate,
and evaluated its lifetime from the decay to diphoton.
We have placed the constraint on the scalaron mass,
and obtained the constraint on the parameter $\beta$ in the Starobinsky model. 
We have shown that $\beta$ should be extremely small for the scalaron to be a DM
although it is desired to be $\mathcal{O}(1)$ for the modified gravity to be a solution for the DE.
This discrepancy naively implies an answer to the question raised in the abstract:
the scalaron DM scenario may be incompatible with the DE problem.
 
The discrepancy stems from contradiction of two statements: ``the scalaron is heavy due to the chameleon mechanism," and ``the schalaron should be light to be a DM candidate."
In $F(R)$ gravity for the cosmology, 
the chameleon mechanism is designed to make the scalaron extremely heavy in the high-density region
in order to avoid the Solar-System constraint.
If one uses the $F(R)$ gravity only for the DE problem, the heavier scalaron is better for the consistency with the constraint.
However, in our scenario, the scalaron should not be too heavy 
because we expect that the scalaron plays the role of a DM candidate.
Thus, the chameleon mechanism or the environment-dependence is the origin of discrepancy 
although it is one of essences in our scenario.

However, we are still left with open questions to avoid the incompatibility.
We should be even skeptical about our result
because some assumptions were made to derive the constraint on the parameter $\beta$.
In order to resolve the discrepancy or incompatibility, 
we need to revisit the assumptions and consider new methodology or model.
Hereafter, we shall reconsider the prescriptions in the analysis, and discuss the possible three ways to improve the gigantic suppression factor for the parameter $\beta$.

First, we shall consider the validity of the particle picture of the scalaron.
We have studied the microscopic nature of the scalaron field by using the quantum field theory,
and derived the constraint on the modified gravity through the upper limit of the scalaron mass.
Here, the result implies that the scalaron mass should be extremely heavy 
if we require $\beta = \mathcal{O}(1)$.
In this case, the scalaron cannot be handled with the scalaron field as a particle,
thus, it should be treated in classical way.
If the scalaron is a classical object, 
we need other prescriptions to discuss the stability of scalaron,
and one can expect it is stable and does not decay.

Second, we shall reconsider the energy-momentum tensor of the environment system surrounding the scalaron.
When we estimate the scalaron mass,
the choice of energy-momentum tensor plays a crucial role 
because the effective potential is sensitive to the environment according to the chameleon mechanism.
In our analysis, we assumed that background matter field is expressed as a perfect fluid,
which is the coarse-grained picture of the matter field.
Here, we assume the realistic matters consist of particles,
that is, an atom or a molecule.
If the scalaron is heavy, its Compton wavelength is short,
and the scalaron cannot touch the particles.
On the other hand, the environment between the particles is almost like a vacuum, 
and the Compton wavelength becomes large
because the chameleon mechanism no longer works.
Thus, we may expect that
the Compton wavelength of the scalaron should be comparable with the coarse-graining scale,
that is, interatomic distance,
which means the scalaron cannot be heavy.

Third, we shall discuss the model of $F(R)$ gravity.
If we admit the constraint on the parameter $\beta$, 
we need to construct a new model of $F(R)$ gravity 
where the scalaron can be light enough to be DM even in the high-density environment.
From the viewpoint of cosmology, 
the fifth force should be suppressed and the heavier scalaron is preferred.
On the other hand, the scalaron cannot be too heavy for a DM candidate.
Here, we may tune the form of $F(R)$ function to weaken the effect of chameleon mechanism;
therefore, the scalaron can be light in such a model.
If we can improve the models of $F(R)$ gravity for the DE and the DM,
the compatibility to observations and experiments for the fifth force 
gives a constraint on the $F(R)$ gravity.

Because of the above reasons, 
it may be premature to conclude that the scalaron cannot be a DM candidate.
These open questions will be treated in the future works.
Besides, we will also study and improve the methods to evaluate the thermal history and relic abundance of the scalaron \cite{Gorbunov:2010bn,Gorbunov:2012ij}. 
Our analysis in this paper can be applied for other modified gravity theories, 
which would give more various DM candidates.

Finally, we note the equivalence between the Jordan frame and the Einstein frame.
The Starobinsky model for the late-time acceleration restores the GR in the large curvature limit.
Then, the scalaron can be handled in the perturbative manner;
the operators of scalaron are added to the SM sector perturbatively.
In this paper, we calculated the dimensionful observables in the Einstein frame
and assumed that the constraint to the $F(R)$ in the Einstein frame is equivalent to that in the Jordan frame.
When the effect of scalaron dynamics is assumed to be small, the above observation is valid.
In the small curvature regime, however, we need to consider the scalaron dynamics in non-perturbative way,
then, the equivalence between two frames are still mysterious.

\section{Acknowledgements}
We are deeply grateful to Shin'ichi Nojiri for his constructive advice and useful comments.
T.K. also thanks Sergei D. Odintsov and Emilio Elizalde for fruitful discussions.
This research is supported 
by the Grant-in-Aid for JSPS Fellows \#15J06973 (T.K.),
and by the JSPS Grant-in-Aid for Young Scientists (B) \#15K17645 (S.M.). 

\appendix

\section{The induced scale anomaly} 
\label{appendix:1}

In this appendix, we calculate the scale anomaly from the path integral measure after the field redefinition of fermion field $\psi$.
Then, we see that the couplings between the scalaron and vector fields through the field strength show up.

We start from the SM-fermion kinetic term in the Einstein frame 
coupled to the scalaron $\varphi$ (see also, Eq.~(\ref{FermiWeyl})), 
\begin{align} 
\mathcal{L}_{F} 
=& e^{- 3 \sqrt{\frac{1}{6}} \kappa \varphi} 
\left( \bar{\psi} \tilde{\gamma}^{\mu} D_{\mu} - \frac{3}{2}i \sqrt{\frac{1}{6}} \kappa 
\left( \partial^{\mu} \varphi \right)  \bar{\psi} \tilde{\gamma}_{\mu} \psi \right) 
%\nonumber \\ 
%&& 
%- e^{- 4 \sqrt{\frac{1}{6}} \kappa \varphi}  \left( 
%m_F \bar{\psi} \psi \right) 
\,, \label{Lag:1}
\end{align}
where we have omitted the spin connection term for simplicity, 
and $D_\mu \psi = \partial_\mu \psi - i g A_\mu \psi$ 
is a covariant derivative regarding an $SU(N)$ gauge symmetry with 
the gauge field $A_\mu = A_\mu^a T^a$, the gauge coupling $g$
 and the generator $T^a$ ($a=1,\cdots , N^2-1$) 
normalized as ${\rm tr}[T^aT^b]=\delta^{ab}/2$. 
The non-minimal coupling, the second term in line one of Eq.(\ref{Lag:1}), 
can be eliminated by making a scale transformation for the fermion field $\psi$ 
along with the scalaron field: 
\begin{equation} 
 \psi \to \psi' = e^{\frac{3}{2} \sqrt{\frac{1}{6}} \kappa \varphi} \psi 
\,, \label{trans}
\end{equation} 
so that one finds 
\begin{align} 
\mathcal{L}_{F} 
\rightarrow 
\mathcal{L}_{F'} 
= \bar{\psi}' \tilde{\gamma}^{\mu} D_{\mu} \psi' 
%- e^{- \sqrt{\frac{1}{6}} \kappa \varphi}  
%m_F \bar{\psi}' \psi' 
\,. \label{Lag:2}
\end{align} 
Thus, the fermion-kinetic term 
in the Einstein frame appears to 
 have no coupling to the scalaron. 
This implies a scale symmetry for the fermion-kinetic term 
associated with the transformation in Eq.(\ref{trans}).

Actually, however, the scale symmetry corresponding 
to the transformation in Eq.(\ref{trans}) 
turns out to be anomalous, so one should have the scale anomaly.  
When one quantizes the SM sector by the path integral formalism,  
this anomaly can be seen from the Jacobian arising from the 
field redefinition by Eq.(\ref{trans}), 
so that one finds the scale-anomaly induced couplings 
between the scalaron and the SM gauge field.

To demonstrate the presence of the scale anomaly, 
we shall first expand the fermion and anti-fermion fields, 
$\psi$ and $\bar{\psi}$, on the basis of the mass eigenstates 
$\{ \psi_n \}$ and $\{ \hat{\psi} \}$ as 
\begin{equation} 
\psi(x) = \sum_n a_n \psi_n(x) 
\,, \qquad 
\bar{\psi}(x) = \sum_{n} \hat{a}_{n} \hat{\psi}_{n} 
\,, \label{expand}
\end{equation}
where the Grassmann numbers $a_n$ and $\hat{a}_n$ 
satisfy $\{a_n, \hat{a}_m \} = 0$, and the mass eigenstates 
$\{ \psi_n \}$ and $\{ \hat{\psi} \}$ have the mass eigenvalues 
$\{\lambda_n \}$ for the Dirac operator$(i \hat{\gamma}^\mu D_\mu)$ 
in such a way that
 $(i \hat{\gamma}^\mu D_\mu) \psi_n = \lambda_n \psi_n$ 
and $\hat{\psi}_n (i \hat{\gamma}^\mu D_\mu) = \lambda_n \hat{\psi}_n$. 
Since we are interested in the weak coupling limit, 
$\kappa \varphi \ll 1$, the transformation in Eq.(\ref{trans}) 
can be regarded as an infinitesimal shift of the fermion field 
involving the infinitesimal-scalaron field dependent 
parameter, i.e., 
\begin{equation}
\psi^{\prime}(x) = (1 + \phi(x) ) \psi(x) 
\,, \qquad 
\phi(x) \equiv \frac{3}{2} \sqrt{\frac{1}{6}} \kappa \varphi(x)
\,. 
\end{equation}
 According to this infinitesimal transformation, 
the expansion coefficients $a_n$ and $\hat{a}_n$ in Eq.(\ref{expand}) 
get the infinitesimal shift: 
\begin{align} 
a_{n} \rightarrow
& \, a_{n^{\prime}} 
= \sum_{m} (\delta_{nm} + C_{nm}) a_{m} 
\,, \nonumber \\  
\hat{a}_{n} \rightarrow
& \, \hat{a}_n^{\prime} = \sum_{m} (\delta_{nm} + C_{nm}) \hat{a}_{m} 
\,, \nonumber \\ 
{\rm with} 
\qquad 
C_{mn} =& 
\int d^{4} x \psi^{\dag}_{m} (x) \phi(x) \psi_{n}(x) 
\,, \label{shift:a}
\end{align}
where we used the orthogonality condition for $\{ \psi_n \}$ 
and $\{ \hat{\psi}_n \}$,
 $\int d^4 x \psi^\dag_m(x) \psi_n(x) = \delta_{mn} $. 
 Then the path integral measures, $\Pi_n d a_n d \hat{a}_n$, 
for $\psi$ and $\bar{\psi}$ 
are transformed like 
\begin{align} 
\Pi_{n} d a_{n} d \hat{a}_{n} 
\rightarrow  
\Pi_{n} d a^{\prime}_{n} d \hat{a}^{\prime}_{n} \cdot {\cal J}^{-2} 
\,, \nonumber \\ 
{\rm with} 
\qquad {\cal J} = {\rm det}(1 + C) 
\,. \label{J1}
\end{align}
Using the identity ${\rm det}(1+C)=e^{{\rm tr}{\rm ln}(1+C)}$ 
and expanding it in powers of the infinitesimal $C$, 
one evaluates the Jacobian ${\cal J}$ in Eq.(\ref{J1})
to the leading oder of $C$ to write 
\begin{equation} 
{\cal J} = {\rm exp}[\sum_{n} C_{nn} + {\cal O}(C^2)]
\,. \label{J2} 
\end{equation}

The Jacobian ${\cal J}$ in Eq.(\ref{J2}) 
includes divergences arising from the 
infinite sum of the fermion eigenvalues  
in $\sum_{n} C_{nn}$, so needs to be properly regularized 
in such a way that 
\begin{equation} 
 \sum_n C_{nn} = 
\lim_{M \to \infty} 
\sum_n \int d^4 x \psi^\dag_n(x) \phi(x) \psi_n(x) e^{\lambda_n^2/M^2} 
\,,  
\end{equation} 
with the cutoff scale $M$. 
Noting $(i \hat{\gamma}^\mu D_\mu)^2 \psi_n = \lambda_n^2 \psi_n$ 
and using $(i \hat{\gamma}^\mu D_\mu)^2 = - D^2 + \frac{g}{2} \sigma_{\mu\nu} F^{\mu\nu}$ 
where $D^2=D_\mu D^\mu$ and $\sigma_{\mu\nu}= \frac{i}{2}[\tilde{\gamma}_\mu, \tilde{\gamma}_\nu]$, 
one can calculate the right hand side as follows: 
\begin{align} 
&\sum_{n} C_{nn}
\nonumber \\ 
&=
\int d^4 x \phi(x) 
\nonumber \\
& \qquad
\lim_{M \to \infty}  
\langle x|  {\rm tr} \left[ e^{-\partial^2/M^2}   
\left(1 + \frac{1}{2} \left( \frac{g}{2} \sigma_{\mu\nu} F^{\mu\nu}\right)^2 
\right) \right] 
|x \rangle  
 \nonumber \\ 
&= 
i \int d^4 x \phi(x)  
\lim_{M \to \infty}  
{\rm tr} \left[ 
\frac{M^4}{(4\pi)^2} + \frac{g^2}{4(4\pi)^2} F_{\mu\nu}^2
\right] 
\,, \nonumber \\ 
\end{align}
where the trace acts on the $SU(N)$ gauge and fermion favors. 
Eliminating the vacuum energy term $\propto M^4$, 
one thus finds the Jacobian 
\begin{equation} 
 {\cal J} 
 = {\rm exp}\left[ 
i \int d^4 x \phi(x) 
\cdot 
\frac{g^2}{4(4\pi)^2} {\rm tr}[F_{\mu\nu}^2]
\right] 
\,, 
\end{equation}  
and hence the scale anomaly term, 
\begin{equation} 
{\cal L}_{\rm anomaly} 
=
- \frac{g^2}{2(4\pi)^2} \phi \, {\rm tr}[F_{\mu\nu}^2]
\,. \label{Lag:anomaly}
\end{equation}

For the diphoton $(AA)$ and digluon $(GG)$ couplings to the scalaron, 
by taking into account the gauge charges of three generation quarks and leptons in the SM, 
Eq.(\ref{Lag:anomaly}) reads 
\begin{align} 
\mathcal{L}_{\rm anomaly} 
=& 
\frac{\varphi}{f_\varphi} \cdot 
\left( \frac{\beta^{\rm eff}(e)}{2e} A_{\mu\nu}^2 
+ \frac{\beta^{\rm eff}(g_s)}{g_s} {\rm tr}[G_{\mu\nu}^2] 
\right) 
\,, \nonumber \\ 
\beta^{\rm eff}(e) 
=& 
\frac{e^3}{(4\pi)^2} b^{\rm eff}(e)
\,, \quad 
b^{\rm eff}(e) = - \frac{11 N_g}{2} 
\,, \nonumber \\ 
\beta^{\rm eff}(g_s) 
=& 
\frac{g_s^3}{(4\pi)^2} b^{\rm eff}(g_s) 
\,, \quad 
b^{\rm eff}(g_s) = - \frac{3 N_g}{2}
\,, \nonumber \\ 
f_{\varphi} =& \frac{\sqrt{6}}{\kappa} 
\,, \label{Lag:anomaly:SM}
\end{align}
where $N_g$ is the number of the generations, $N_g=3$,  
$e$ and $g_s$ denote the electromagnetic and QCD gauge couplings, respectively.

\section{The relevant partial decay widths} 
\label{appendix:2} 

In this appendix, we show the calculation of the decay width of scalaron.
The scalaron couplings to the SM fermions ($F$) are given by 
\begin{equation}  
\mathcal{L}_{\varphi FF} 
= \frac{\varphi}{f_{\varphi}} \sum_F m_F \bar{\psi}_{F}^{\prime} \psi_{F}^{\prime} 
\, 
\label{Yukawa}
\end{equation}
(see Eq.~(\ref{FermionWeyl4})).
From this Lagrangian, one can readily compute the partial decay rates to the SM fermions to get 
\begin{equation}  
\Gamma(\varphi \to F\bar{F}) 
= \frac{N_c^{(F)} m^{2}_{F}}{8 \pi f_\varphi^2} m_{\varphi}  \left( 1 - \frac{4 m_F^2}{m_\varphi^2} \right)^{3/2}
\,, \label{decay:FF}
\end{equation}
where $N_{c}^{(F)}=1(3)$ for leptons (quarks).

\begin{figure}[htbp]
\centering
\includegraphics[width=0.5\linewidth]{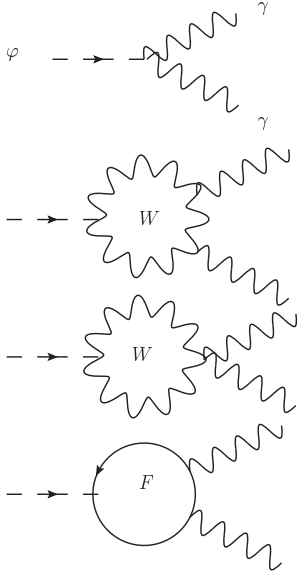} 
\caption{ 
The Feynman graphs relevant to the $\varphi \to \gamma\gamma$ decay processes. 
The W boson loop graphs have been drawn in the unitary gauge.}  
\label{fig-diphoton}
\end{figure}

\begin{figure}[htbp]
\centering
\includegraphics[width=0.5\linewidth]{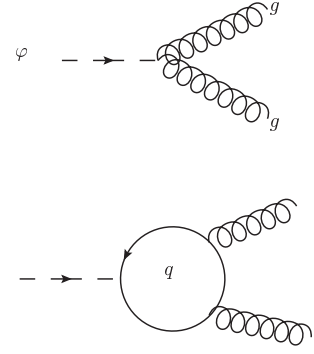} 
\caption{ 
The Feynman graphs relevant to the $\varphi \to gg$ decay processes. } 
\label{fig-digluon}
\end{figure}

As to the couplings to diphoton and digluon, 
in addition to the scale anomaly term in Eq.(\ref{Lag:anomaly:SM}), 
the scalaron couplings are generated at one-loop level of the 
SM perturbation theory through the Yukawa vertices in Eq.(\ref{Yukawa}) and  
the vertices involving the $W$ boson, 
\begin{equation} 
\mathcal{L}_{\varphi WW} 
=  \frac{2 \varphi}{f_\varphi} m_W^2 W_\mu^+ W^{\mu-} 
\,
\end{equation}
(see Eq.~(\ref {VectorWeyl25})).
The net contributions to the partial decay rates are then computed to be 
(See Figs.~\ref{fig-diphoton} and ~\ref{fig-digluon})


\begin{align} 
\Gamma[\varphi \rightarrow \gamma\gamma] 
=& \frac{\alpha_{\rm em}^2 m_\varphi^3}{64 \pi^3 f_\varphi^2} 
\left | b^{\rm eff}(e) + a_W(\tau_W)  
\right.
\nonumber \\ 
& \left. \qquad \qquad 
+ \sum_{F} N_{c}^{(F)} Q_F^2 a_F(\tau_F)  \right |^2
\,, \nonumber \\ 
\Gamma[\varphi \rightarrow gg] 
=& \frac{\alpha_{s}^2 m_\varphi^3}{32 \pi^3 f_\varphi^2} 
\left| b^{\rm eff}(g_s) + \sum_{F={\rm quarks}} a_F(\tau_F)  \right|^2
\,,  \label{widths:2}
\end{align}
where $\alpha_{\rm em}=e^2/(4\pi), \alpha_s = g_s^2/(4\pi)$ and  
\begin{align} 
a_{W}(\tau_{W}) 
=&
- \frac{1}{2} \left[ 2 + 3 \tau_W + 3 \tau_W (2- \tau_W) f(\tau_W) \right]
\,, \nonumber \\ 
a_{F}(\tau_F) 
=&
\tau_F \left[ 1 + (1- \tau_F) f(\tau_F) \right]
\,, \nonumber \\ 
f(\tau) 
=&  \left \{ 
\begin{array}{cc} 
\left[ \sin^{-1}\left( \frac{1}{\sqrt{\tau}}  \right) \right]^2 & {\rm for} \qquad \tau >1 \\ 
- \frac{1}{4}\left[ {\rm ln} \left(  \frac{1 + \sqrt{1+\tau}}{1 - \sqrt{1- \tau}} \right) - i \pi \right]^2 
& {\rm for} \qquad \tau \le 1
\end{array}
\right.  
\,, \nonumber \\ 
\tau_{W/F} =& \frac{4 m_{W/F}^2}{m_\varphi^2}   
\,. 
\end{align}

\end{document}